\documentclass[aps,eqsecnum,preprint,preprintnumbers,12pt,amsfonts]{revtex4}

\usepackage{epsfig,psfrag}
\usepackage{latexsym}
\usepackage{pictex}

\newcommand{\beq}{\begin{eqnarray}}
\newcommand{\eeq}{\end{eqnarray}}
\newcommand{\nn}{\nonumber}

\newcommand{\reals}{\mbox{${\rm I\!R }$}}
\newcommand{\sumn}{\sum_{n=1}^\infty}

\newcommand{\sumdp}%
{{\sum_{n_1,...,n_d=-\infty}^\infty}^{\hspace{-.5cm}\prime\hspace{.5cm}}}

\usepackage{color}

\begin{document}

\title{Comment on: ``The Casimir force on a piston in the
spacetime with extra compactified dimensions''
[Phys.~Lett.~B 668 (2008) 72]}
\author{S. A. Fulling}\email{fulling@math.tamu.edu}
\affiliation{Departments of Mathematics and Physics, Texas A\&M University, College Station, TX, 77843-3368 USA}
\author{Klaus Kirsten}\email{klaus_kirsten@baylor.edu}
\affiliation{Department of Mathematics, Baylor University, Waco,
TX 76798-7328 USA}
\date{\today}

\begin{abstract}
We  offer a clarification of the significance of the indicated paper of
H. Cheng. Cheng's conclusions about the attractive nature of Casimir forces between parallel plates are valid beyond the particular
model in which he derived them; they are likely to be relevant to other
recent literature on the effects of hidden dimensions on Casimir forces.
\end{abstract}

\maketitle

The interesting Letter by Cheng  \cite{Cheng08}
studies the effect of small Kaluza--Klein dimensions on the Casimir
force between parallel plates in the macroscopic dimensions.
It reaches conclusions about parallel plates
different from those of a previous paper by the
same author \cite{Cheng06}.
Here we offer some remarks that we believe make its significance
clearer.

\begin{figure}
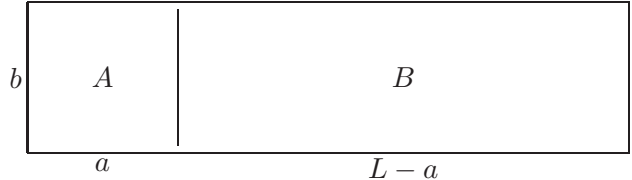

\centerline{\beginpicture
 \setcoordinatesystem units <2truecm,2truecm>
 \putrule from 0 0 to 4 0
 \putrule from 0 1 to 4 1
 \putrule from 0 0 to 0 1
 \putrule from 1 0.05 to 1 0.95
 \putrule from 4 0 to 4 1
 \put{$b$} [r] <-2pt,0pt> at 0 0.5
  \put{$a$} [t] <0pt,-2pt> at 0.5 0
   \put{$L-a$} [t] <0pt,-2pt> at 2.5 0
\put{$A$} at 0.5 0.5
\put{$B$} at 2.5 0.5
\endpicture}
\caption{A schematic Casimir piston.
The interior plate is free to move.}
\label{fig:piston} \end{figure}

First, a historical observation.
In early work on the Casimir force between parallel plates it was
standard to enclose the movable plate in a large, finite box, so
as to make all energies finite.  This procedure is followed, for
example, in \cite{Fierz,Power,Boyer}.
  That is, all these early calculations were done in what we would
nowadays call a piston \cite{Cav,HJKS}.
At the end the transverse dimensions
(such as $b$ in Fig.~\ref{fig:piston}) and the length $L-a$ in
Fig.~\ref{fig:piston} were taken to
infinity; it was found that the results then were the same as
those of a more naive calculation where the transverse boundaries
and the distant plate (right end of Fig.~\ref{fig:piston}) were never
present.
Emboldened by this consistency, many later authors did not bother
to put in the spatial cutoffs.

 Indeed, that looser procedure was followed by Cheng in \cite{Cheng06}
but not in \cite{Cheng08}.   The earlier paper concluded that the
presence of compactified extra dimensions modified the Casimir force,
making it repulsive under certain conditions.
(In terms of Fig.~\ref{fig:piston}, the narrow chamber $A$ would tend
to expand, not contract, in analogy with the well known (but heavily
disputed) prediction for a conducting rectangular cavity \cite{Luk}.)
However, the more recent paper \cite{Cheng08} studies a fully finite
configuration like that of Fig.~\ref{fig:piston} (but with extra
dimensions, both small and large).  One finds that the
result of moving the rightmost plate to infinity is \emph{not} the same
as never having that plate at all.  The force is always attractive.

\goodbreak

This result is  analogous to that in the original piston
literature \cite{Cav,HJKS}, where careful calculation of the
total vacuum energy in the  rectangular cavities $A$ and $B$ yields an
attractive force, notwithstanding~\cite{Luk}.
The details of this analogy are worth dwelling upon.
The point of \cite{Cav,HJKS} (see also \cite{rect}) is that
 regarding chamber $B$ as
infinitely long (or as absent) obscures the fact that a change in the
energy in
$B$ cancels a part of the change in the energy in $A$ when the piston
plate moves.
On the other hand, the existence of the finite transverse dimension(s)
$b$ is essential for creating the repulsive force in $A$ to begin with.
That force arises from the change with $a$ of the (negative) Casimir
energy of interaction of the top and bottom plates in
Fig.~\ref{fig:piston}.
In the calculations of Cheng the large transverse dimensions play no
essential role in the considered limit $b\to\infty$.
However, the small extra dimensions are now significant.
The repulsive force found in Cheng's first paper \cite{Cheng06}
is (minus) the $a$-derivative of the Casimir energy associated with the
Kaluza--Klein dimensions, only cavity $A$ being considered.
In his second paper \cite{Cheng08} the Kaluza-Klein vacuum energy
is taken into account also for cavity $B$, so that the repulsive effect
is cancelled.

What makes the model in \cite{Cheng08} a ``piston'' is not just the
presence of the third plate, but also the presence of compact transverse
dimensions.  Both elements are necessary, one to create the ``paradox''
of a repulsive force and the other to remove it.
 In the Kaluza-Klein context this situation was not obvious until Cheng
pointed it out; see also \cite{EMN}.
As he stresses, experimental evidence is against
any repulsion between large parallel plates;
although originally offered \cite{Cheng06}  as evidence against the
existence of Kaluza-Klein dimensions,  this fact now \cite{Cheng08}
appears as additional confirmation that a careful ``piston'' analysis is
the correct way to do the parallel-plate calculation if such dimensions
do exist.

In closing we make a technical comment on Cheng's calculations.
(We note that more recent treatments \cite{LT,EM} of multidimensional
pistons are more general than previous literature.)
Both papers \cite{Cheng06,Cheng08} are based on a mode sum that
corresponds, in the small compact dimensions, to the spectrum of a
rectangular box with Neumann boundary conditions.  It would be more
natural to take periodic boundary conditions (i.e., to make the
Kaluza--Klein space  a torus) or to consider non-flat compact extra dimensions like a sphere.
But in fact, the main conclusions of \cite{Cheng06,Cheng08}
are independent of such details.
 Let $N$ denote the manifold of the extra dimensions such that the space
of our universe is given by $\reals^3 \times N$, the piston living in
$\reals^3$. In the limit where $b\to \infty$ in Fig. 1, the
eigenfrequencies of the vacuum fluctuations are given by
$$\omega^2 =
k_1^2+k_2^2+\left( \frac{n\pi} D\right) ^2 +\lambda_i^2, $$
where $n$ and $i$ are positive integers, $k_1^2+k_2^2$ comes from the
two free
transversal dimensions in $\reals^3$, $(n\pi/D)^2$ results from the
Dirichlet plates ($D=a$ for the left chamber and $D=L-a$ for the right
chamber in Fig. 1), and $\lambda_i^2$ are the eigenfrequencies in the
additional dimensions,
$$-\Delta_N \varphi_i= \lambda_i^2 \varphi_i.$$
Under the assumption that $\lambda_i^2 \geq 0$, the Casimir force on the
piston, as $L\to\infty$, can be shown to be
\beq F&=& - \frac{\pi^2
g_0}{480 a^4}   + \frac 1 {8\pi^2} \sumn {\sum_i}'
\frac{\lambda_i^2}{n^2} \frac \partial {\partial a} \frac 1 a K_2 (2an
\lambda_i),\nn\eeq
where $g_0$ is the multiplicity of the states with
$\lambda_i=0$ and the prime indicates that the summation over $i$ omits these states.
Arguing as in \cite{Cav}, using well known properties of
the modified Bessel function $K_\nu (z)$ \cite{grad65b}, $F$ can be
seen to be negative (as in \cite{Cheng08})
independently of any details of the topology or
geometry of the extra dimensions (within the confines $\lambda_i^2 \geq
0$). Note that this conclusion remains valid if we replace Dirichlet by Neumann
boundary conditions on the plates, which implies we sum $n$ from zero to infinity.
This only introduces additional $D$-independent terms into the energy which are irrelevant
for the force.

With only the left chamber in Fig. 1 considered,
 the answer in general
will contain renormalization ambiguities.
However, on the equilateral torus of radius $R$ the force is finite and reads
\beq F &=& - \frac{\pi^2} {480 a^4}
+ \frac{\Gamma \left( \frac d 2 +2\right)}
{32 \pi^{4+d/2} R^4} Z_d \left(\frac d 2 +2\right) \nn\\
& &+\frac 1 {8\pi^2 R^2} \sumn \sumdp \frac{n_1^2+...+n_d^2}{n^2}
 \frac \partial {\partial a}
\frac 1 a K_2 \left( \frac{2an} R \sqrt{n_1^2+...+
n_d^2}\right),\nn\eeq
where $$Z_d (s) = \sumdp \left[ n_1^2+...+n_d^2\right]^{-s}.$$
This shows  (as in \cite{Cheng06})
that for $a \gg R$ the force is positive
and asymptotically constant when region $B$ and its spatial cutoffs are
neglected.
%stressing the point once again that not putting spatial cutoffs might
%give wrong results.

The only effect of
replacing Neumann by periodic conditions in the extra dimensions
is to multiply the contribution of each
nonzero quantum number by $2$, since it corresponds to both a
left-moving and a right-moving mode.
Of course, one does not know what the actual geometry of the
Kaluza-Klein space is.
The arguments we have just presented
(whose details will be presented elsewhere \cite{inprep})
show that the situation is qualitatively similar for
any such geometry, with the caveat that the
(presumably spurious) Lukosz-type repulsion for a single chamber will
not arise at all unless the (renormalized) Casimir energy of the
Kaluza--Klein manifold is negative, as it is in a torus of any
dimension.
There are cases where that energy is positive --- for example, a scalar
field in a 2-dimensional rectangle with Dirichlet conditions or an
electromagnetic field in a 3-dimensional rectangular box with perfect
conductor conditions \cite{AW}.
Note that this question is distinct from (although related to) the issue
of whether a Lukosz repulsion exists within the Kaluza--Klein space
itself:  positive energy is not synonymous with positive
force.

%\begin{acknowledgement}
\subsection*{Acknowledgments}
Our work is supported by National Science Foundation Grants PHY-0554849
(TAMU) and PHY-0757791 (Baylor). S.A.F. enjoyed the hospitality of the Institute for Mathematics
and Its Applications (U. Minnesota, Minneapolis) while the paper was written.
%\end{acknowledgement}

\bibliographystyle{elsarticle-num-names}

\end{document}